\pdfoutput=1
\documentclass[aps,prl,10pt,twocolumn,superscriptaddress,nofootinbib,floatfix,amsmath,amssymb]{revtex4-2}
\usepackage{graphicx}
\usepackage{hyperref}
\hypersetup{colorlinks=true,linkcolor=blue,citecolor=blue,urlcolor=blue,
  pdftitle={Gauge Field Fluctuations Protect the Higgs Vacuum When a Small Universe Begins}}
\usepackage{orcidlink}
\newcommand{\runin}[1]{\textit{#1}---}%
\newcommand{\PbarOne}{0.73}        %
\newcommand{\PbarOneAbs}{0.7}      %
\newcommand{\PbarSigma}{0.86}      %
\newcommand{\PbarInterp}{0.68}     %
\newcommand{\PbarBKPV}{0.63\text{--}0.67} %
\newcommand{\BayesMax}{0.75}       %
\newcommand{\mExclFCC}{173.7}      %
\newcommand{\sigFCC}{5.3}          %
\newcommand{\sigHLLHCmin}{1.7}     %
\newcommand{\sigHLLHCmax}{2.5}     %
\newcommand{\xiHmin}{0.08}         %
\newcommand{\xiHcentral}{0.074}    %
\newcommand{\xiHhalf}{0.12}        %
\newcommand{\marginGeV}{0.36}      %
\newcommand{\marginLam}{0.0024}    %
\newcommand{\mcMinusMzero}{1.9}    %
\newcommand{\modelErrUp}{0.07}     %
\newcommand{\modelErrDn}{0.08}     %
\newcommand{\modelErrUpNoVi}{0.03}      %
\newcommand{\modelErrDnNoVi}{0.05}      %
\newcommand{\xiHminEW}{0.09}            %
\newcommand{\xiHbandSM}{0.013}          %
\newcommand{\xiHbandAs}{0.04}           %
\newcommand{\mSurvFCC}{172.7}         %
\newcommand{\sigFCCamb}{6.0}          %
\newcommand{\sigHLLHCinterpMin}{1.1}  %
\newcommand{\sigHLLHCinterpMax}{1.2}  %
\newcommand{\lamcErrUp}{0.0006}       %
\newcommand{\lamcErrDn}{0.0005}       %
\newcommand{\PoneAtExclFCC}{0.023} %
\newcommand{\BexFCC}{0.033}        %
\providecommand{\aFitA}{5.915}%
\providecommand{\bFitA}{1.306}%
\providecommand{\resFitA}{0.06}%
\providecommand{\lamFitLo}{-0.018}%
\providecommand{\lamFitHi}{-0.007}%
\providecommand{\sigEffA}{0.44}%
\providecommand{\closedAccSmall}{1.1\%}%
\providecommand{\closedAccLarge}{3.2\%}%
\providecommand{\lcAhundred}{-0.0144}%
\providecommand{\lcAthree}{-0.0118}%
\providecommand{\lcAtenk}{-0.0072}%
\providecommand{\lcDthree}{-0.0115}%
\providecommand{\lcEthree}{-0.0128}%
\providecommand{\lamTop}{-0.028}%
\providecommand{\lcAthreeErrUp}{0.0003}%
\providecommand{\lcAthreeErrDn}{0.0002}%
\providecommand{\topGapLo}{0.3}%
\providecommand{\topGapHi}{0.9}%
\providecommand{\SwkbLo}{2}%
\providecommand{\SwkbHi}{8}%
\providecommand{\NeffHalf}{250}%
\providecommand{\NstarA}{970}%
\providecommand{\NstarD}{590}%
\providecommand{\NstarE}{1.8\times10^5}%
\providecommand{\zpSlopeLo}{0.7}%
\providecommand{\zpSlopeHi}{1.2}%
\providecommand{\dVredLo}{12\%}%
\providecommand{\dVredHi}{32\%}%
\providecommand{\crossShiftLc}{0.0008}%
\providecommand{\SzeroGeneric}{10^{-11}}%
\providecommand{\staticHazardDiff}{0.0003\text{--}0.0004}%
\providecommand{\kappaZero}{0.48}%
\providecommand{\kappaMin}{0.43}%
\providecommand{\lcSthirtyGtwo}{22\%}%
\providecommand{\lcSthirtyGY}{6\%}%
\providecommand{\lcSthirtyYt}{18\%}%
\providecommand{\abelLamMin}{-1}%
\providecommand{\abelLamMax}{-0.0065}%
\providecommand{\abelSlopeLo}{1.8}%
\providecommand{\abelSlopeHi}{4.0}%
\providecommand{\abelSlopeWKB}{4.44}%
\providecommand{\abelBOdev}{30\%}%
\providecommand{\abelCAP}{10^{-6}}%
\providecommand{\abelLamcSM}{-0.071}%
\providecommand{\abelHolonomy}{9\times10^{-4}}%
\newcommand{\MPbar}{\bar M_{\mathrm{P}}}         %

\newcommand{\NC}{N_C}                            %

\newcommand{\MSbar}{\overline{\mathrm{MS}}}
\newcommand{\mtbar}{\bar m_t}                    %
\newcommand{\mub}{\mu_b}                         %
\newcommand{\Lb}{L_b}                            %
\newcommand{\lamc}{\lambda_c}                    %
\newcommand{\xiH}{\xi_H}                         %
\newcommand{\xic}{\xi_c}                         %
\newcommand{\rT}{r_{\mathrm{T}}}                 %

\graphicspath{{figures/}}
\begin{document}

\title{Gauge Field Fluctuations Protect the Higgs Vacuum When a Small Universe Begins}
\author{Alan Zhang\,\orcidlink{0009-0000-1187-0755}\,\textsuperscript{$*$}}
\affiliation{Department of Physics and Astronomy, McMaster University, Hamilton, Ontario L8S 4M1, Canada}
\affiliation{Perimeter Institute for Theoretical Physics, Waterloo, Ontario N2L 2Y5, Canada}

\begin{abstract}
The Standard Model Higgs quartic coupling $\lambda$ turns negative at high energies.
We show that in a small periodic box with antiperiodic fermions, zero-point fluctuations of the constant gauge modes turn the Higgs zero mode into a resonance for $\lambda<0$, whose width falls exponentially in $|\lambda|^{-1/2}$. A box lasting $N$ light-crossing times keeps its vacuum with probability above $1/2$ for $\lambda>\lambda_c(N)$, which depends on $N$ only logarithmically.
For a universe that begins at a Casimir turning point, $\lambda_c\approx-0.012$ lies below the Standard Model value $-0.0098$; given a curvature coupling $\xi_H\gtrsim\xiHmin$, survival has probability above $1/2$ for $m_t<172.96$~GeV and $\approx\PbarOneAbs$ for current data, and FCC-ee can exclude this beginning if $m_t\gtrsim\mExclFCC$~GeV. %
\end{abstract}

\maketitle
{\renewcommand{\thefootnote}{\fnsymbol{footnote}}\footnotetext[1]{Contact author: \href{mailto:zhana142@McMaster.ca}{zhana142@McMaster.ca}}}

\runin{Introduction}The measured masses of the Higgs boson and the top quark place the Standard Model (SM) near the boundary between a stable and a metastable electroweak vacuum~\cite{Isidori:2001bm,Degrassi:2012ry,Bezrukov:2012sa,Alekhin:2012py,Buttazzo:2013uya,Bednyakov:2015sca,Andreassen:2017rzq,Chigusa:2017dux,Chigusa:2018uuj}.
For central values the Higgs quartic coupling $\lambda$ turns negative above $10^{10}$--$10^{11}$~GeV, and our vacuum survives only because tunneling to large field values is slow~\cite{Coleman:1977py,Buttazzo:2013uya}. %
During and after inflation, curvature and quantum fluctuations can push the Higgs field over its barrier, which bounds the inflationary Hubble rate and the Higgs curvature coupling $\xiH$~\cite{Espinosa:2007qp,Herranen:2014cua,Hook:2014uia,Kearney:2015vba,Herranen:2015ima,Espinosa:2015qea,Markkanen:2018pdo,Mantziris:2020rzh}.
These analyses assume a volume much larger than the scale of the instability.
In a small periodic box of edge $L$, modes with nonzero momentum cost energies of order $2\pi/L$, and the spatially constant modes of the Higgs and gauge fields form a quantum-mechanical system, as in small-volume gauge theory~\cite{Luscher:1982ma,Luscher:1983gm,Koller:1987fq,vanBaal:2000zc}.
For $\lambda(1/L)<0$ the Higgs zero mode alone would run away.

We show that the zero-point fluctuations of the constant gauge modes prevent this.
At a fixed Higgs zero mode these modes are oscillators with frequencies proportional to the Higgs field, so their zero-point energy grows linearly with it.
Such terms close the flat valleys of small-volume Yang--Mills theory and of bosonic matrix models~\cite{Simon:1983jy,vanBaal:2000zc,Barbon:2005zj}.
They are the finite-volume counterpart, with the opposite sign, of the term cubic in the gauge-boson masses that the static Matsubara modes add to the thermal potential~\cite{Dine:1992wr,Arnold:1992rz}. Bosonic modes without a geometric gap give a term non-analytic in $\Phi^\dagger\Phi$~\cite{Barbon:2006us}, absent from the infinite-volume effective potential, and antiperiodic fermions give none.
Here the linear term holds back a negative quartic.
The electroweak state of the box becomes a resonance whose width falls exponentially in $|\lambda|^{-1/2}$, and a box lasting $N$ light-crossing times keeps its vacuum down to a finite threshold $\lamc(N)$, which depends on $N$ only logarithmically.

\begin{figure*}[t]
\centering
\includegraphics[width=\textwidth]{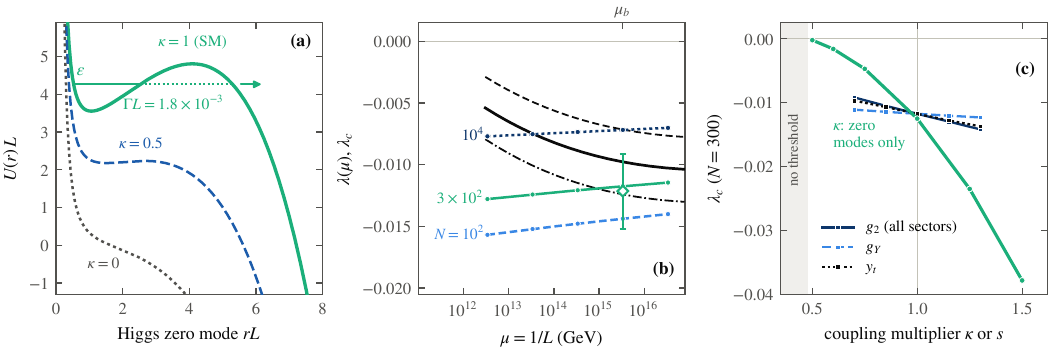}
\caption{The finite-volume threshold in a static box.
(a)~Radial potential of the Higgs zero mode for $\lambda=-0.012$ and SM couplings at $\mu=1/L=3.4\times10^{15}$~GeV (model without cross terms), with the gauge zero-mode couplings scaled by $\kappa$ ($\Delta V$ fixed); for $\kappa=1$ the electroweak state is a resonance at $\varepsilon$ with width $\Gamma$. %
(b)~$\lamc$ ($P=1/2$) for $N$ light-crossing times against $\mu=1/L$ (dots: computed boxes), and SM $\lambda(\mu)$ for $m_t=172.2$ (dashed), 172.6 (solid) and 173.0~GeV (dash-dotted); diamond: expanding torus of a Casimir turning point (bar: $P_1=0.9$--0.1). %
(c)~$\lamc$ for $N=300$ at the scale of (a), with the gauge zero-mode couplings scaled by $\kappa$ [model of (a)] or $g_2$, $g_Y$ or $y_t$ by $s$ (full model).} %
\label{fig:mechanism}
\end{figure*}

A small compact volume speeds up vacuum decay by replacing the bubble with a uniform tunneling mode~\cite{Hassan:2026cmp,Brown:2015kgj}.
For the Higgs field at $\lambda<0$ this mode has no barrier of its own; when space is a three-torus, the gauge zero modes supply one. Neither those analyses nor the thin-wall analysis of decay with a compact dimension~\cite{Antoniadis:2024ent} includes them.
The threshold therefore applies to a universe that begins as a small torus whose fermions have no constant modes, created from nothing~\cite{Zeldovich:1984vk,Coule:1999wg,Linde:2004nz} or turned around by Casimir energy~\cite{Guth:2025dal}. A torus that starts with edge $\MPbar^{-1}\ll L_0\ll1/H$ and then inflates holds the zero mode in the box at $\mu\approx1/L_0$ for $N$ of order $1/(HL_0)$ light-crossing times, its conformal duration, so that Fig.~\ref{fig:mechanism}(b) maps its survival.
At these scales the SM quartic, $\lambda\approx-0.01$, is of the same size as $\lamc(N)$ for $N\sim10^2$--$10^4$, so this effect, together with $\xiH$, decides whether the vacuum survives such a beginning. %
For compactifications to lower dimensions, earlier work studied the SM Casimir energy on circles and tori~\cite{Arkani-Hamed:2007ryu,Fornal:2011tw} and its Wilson lines together with the high-scale Higgs vacuum~\cite{Hamada:2017yji}; to our knowledge the resonance, its threshold and the top-quark criterion that follows from them are new.

In this Letter we compute the resonance for the SM with one-loop finite-volume matching, determine its threshold in a static box as a function of the duration, the size of the box and the couplings [Fig.~\ref{fig:mechanism}], and, with the companion paper~\cite{v12PRD}, apply it to a flat universe that begins at a Casimir turning point, where $\lamc=-0.0121$. %
The SM value clears this threshold whenever $m_t<172.96$~GeV, or $\mtbar<163.16$~GeV for the $\MSbar$ mass $\mtbar\equiv\mtbar(\mtbar)$, a criterion that FCC-ee can test. %

\runin{Zero modes in a small box}Consider the SM on a flat three-torus of edge $L$, without 't~Hooft flux and with fermions antiperiodic around all three cycles, so that the fermions have no constant modes and the one-loop vacuum lies at trivial holonomy~\cite{vanBaal:1988va}.
The constant modes of the Higgs doublet and of the electroweak connections have no tree-level gap; in units of $1/L$ their Hamiltonian is
\begin{multline}
LH=\tfrac12\bigl(p_\Phi^2+\mathbf p_x^2+\mathbf p_w^2\bigr)+\tfrac{g_2^2}{2}\sum_{i<j}(\mathbf x_i\times\mathbf x_j)^2\\
+\sum_i\bigl|(g_2\,\mathbf x_i\cdot\mathbf T+g_YYw_i)\Phi\bigr|^2+\tfrac{\lambda}{4}r^4+\Delta V(r),
\label{eq:H}
\end{multline}
where $\mathbf x_i$ and $w_i$ ($i=1,2,3$) are the constant SU(2) and hypercharge connections in canonical normalization, $\mathbf T$ are the SU(2) generators, $\Phi$ is the Higgs zero mode with $\Phi^\dagger\Phi=r^2/2$, and the couplings are $\MSbar$ couplings at $\mu=1/L$.
Physical states are invariant under constant gauge transformations, and $\Delta V(r)$ is the one-loop energy of the nonzero modes of all SM fields at zero holonomy (End Matter A).

At fixed $r$ the $W$ and $Z$ zero modes are oscillators with frequencies $g_2r/2$ and $g_Zr/2$, where $g_Z=(g_2^2+g_Y^2)^{1/2}$, so their zero-point energy grows linearly in $r$ with slope $\sigma_0=\tfrac32g_2+\tfrac34g_Z\approx1.31$. %
On balance the nonzero modes do not confine: in $\Delta V$ the periodic bosons lower the potential, and the antiperiodic top-quark tower, which raises it, offsets them only partly, so that $\Delta V$ roughly halves the barrier above the resonance level (from $1.57/L$ to $0.72/L$ at $\lambda=-0.010$)~\cite{v12Report}. %
Against the quartic, $U(r)\simeq\sigma r-|\lambda|r^4/4$ has a well and a barrier, and the electroweak ground state becomes a resonance $E=\varepsilon-i\Gamma/2$ [Fig.~\ref{fig:mechanism}(a)].
For a level deep in the well, tunneling through $U$ gives
\begin{equation}
\Gamma\simeq A\exp\Bigl(-\frac{2\pi\sqrt2}{3}\,\frac{\sigma}{|\lambda|^{1/2}}\Bigr).
\label{eq:Gamma}
\end{equation}
Constant fermion modes would act against the confinement: if all three cycles were periodic, as for the torus of Ref.~\cite{Fornal:2026qfq}, the constant modes of the top quark would add $-6m_t\approx-1.8\,r$ to the zero-point energy and outweigh $\sigma_0r$, much as supersymmetry cancels the zero-point confinement of the bosonic membrane~\cite{deWit:1988xki}. %

\runin{Threshold}In a static box that exists for a time $T=NL$, the zero mode is still in the resonance with probability $P=e^{-\Gamma N}$, and we define the finite-volume threshold $\lamc(N)$ as the value of $\lambda(1/L)$ at which $P=1/2$.
We compute $\Gamma(\lambda)$ from Eq.~\eqref{eq:H}, with $g_2$, $g_Y$ and $y_t$ fixed at their SM values at $\mu=1/L$ (End Matter A).
For $\mu=3.4\times10^{15}$~GeV and $\lamFitLo\leq\lambda\leq\lamFitHi$ the widths follow Eq.~\eqref{eq:Gamma} with an effective $\sigma=\sigEffA$, that is, $\ln(\Gamma L)=a-b|\lambda|^{-1/2}$ with $a=\aFitA$ and $b=\bFitA$ (root-mean-square residual $\resFitA$). %
In this window the level lies only $\topGapLo$--$\topGapHi/L$ below the barrier top (WKB exponents $\SwkbLo$--$\SwkbHi$), the zero-point energy rises with slope $\zpSlopeLo$--$\zpSlopeHi$ rather than $\sigma_0$, and, at $\lambda=-0.012$, $\Delta V$ lowers that slope by $\dVredLo$--$\dVredHi$; the finite resonance energy lowers the exponent even in the abelian Higgs model (End Matter A). %
Setting $\Gamma N=\ln2$ gives $\lamc(N)=-[b/(a+\ln(N/\ln2))]^2$, within $\closedAccSmall$ of the computed thresholds for $10^2\leq N\leq10^4$ and $\closedAccLarge$ at $N=10^5$; for example $\lamc=\lcAhundred$, $\lcAthree^{+\lcAthreeErrUp}_{-\lcAthreeErrDn}$ (model error) and $\lcAtenk$ at $N=100$, 300 and $10^4$ [Fig.~\ref{fig:mechanism}(b)]. %
Like the metastability bound, $\lamc$ is defined by a duration, on which it depends only logarithmically ($10\lesssim N\lesssim10^6$). %
The SM vacuum itself keeps $P>1/2$ for $N<\NstarA$ at this scale, and for $N<\NstarD$ to $\NstarE$ as $\mu$ goes from $3.4\times10^{16}$ to $3.4\times10^{12}$~GeV. Over this range $\lamc(300)$ stays between $\lcEthree$ and $\lcDthree$: the size of the box enters only through the couplings at $\mu=1/L$. %
The barrier is due to the gauge zero modes: when their couplings in Eq.~\eqref{eq:H} are scaled by $\kappa$ with $\Delta V$ fixed, the threshold for $N=300$ vanishes at $\kappa\approx\kappaZero$, and at $\kappa=0$ the radial potential falls monotonically for every $\lambda\leq0$ [Fig.~\ref{fig:mechanism}(a),(c)]. %
Scaling $g_2$, $g_Y$ or $y_t$ by a factor between 0.7 and 1.3 moves $|\lamc|$ by at most $\lcSthirtyGtwo$, $\lcSthirtyGY$ or $\lcSthirtyYt$, and increasing any of them deepens the threshold. %

\runin{Accuracy}Three approximations, shared by the static box and the expanding torus below, lead from Eq.~\eqref{eq:H} to $\Gamma$: the one-loop matching, in Coulomb gauge with the $\MSbar$ finite-volume sum at $\mu=1/L$; the reduction of the zero modes to $r$ and the hypercharge modulus $|\mathbf w|$, with the SU(2) modes in a conditional Gaussian corrected by diffusion Monte Carlo (DMC); and the cross terms through which the holonomy dependence of the Casimir energy of every SM field couples the zero modes to the nonzero modes.
We have tested each (End Matter A).
The cross terms are the largest correction and make $\lamc(300)$ shallower by $\crossShiftLc$; together the tests give the model error quoted above, $+\modelErrUpNoVi/{-\modelErrDnNoVi}$~GeV in the critical top mass below. %
Blind reimplementations from a specification of the model reproduce the critical masses without cross terms to 0.005~GeV and the cross terms to 0.01~GeV. %
Inhomogeneous tunneling does not compete: a Fubini-type bounce has action $8\pi^2/(3|\lambda|)\approx2.6\times10^3$ at $|\lambda|=0.01$~\cite{Andreassen:2017rzq}, against zero-mode tunneling exponents of $\SwkbLo$--$\SwkbHi$. %

\runin{A universe that begins small}Turning a contracting flat universe around requires a violation of the null energy condition~\cite{Molina-Paris:1998xmn,Novello:2008ra}.
On a small three-torus with antiperiodic fermions the negative Casimir energy of the SM fields supplies it: with a minimally coupled inflaton on the Starobinsky plateau~\cite{Starobinsky:1980te}, of Hubble rate $h$, the Casimir energy density $\rho_C=-\NC\gamma_C/L^4$ turns a torus of edge $L$ around~\cite{Zeldovich:1984vk,Guth:2025dal,v12PRD},
\begin{equation}
L=\Lb\sqrt{\cosh2ht},\qquad \Lb^4=\frac{\NC^{\rm eff}\gamma_C}{3\MPbar^2h^2}.
\label{eq:turn}
\end{equation}
For the SM, with $\gamma_C=1.675$ and $\NC^{\rm eff}=17.9$ (the value $\NC=25.5$ lowered by the zero modes), this gives $\Lb\simeq722\,\MPbar^{-1}$, a few thousand Planck lengths and $230$ times smaller than $1/h$. %
The turning point is nearly empty, its negative energy paying for only about five quanta, and we read it as a beginning from which both branches of the solution emerge~\cite{v12PRD}. %
The Higgs zero mode starts in the resonance at $\mub=1/\Lb=\MPbar/722\approx3.4\times10^{15}$~GeV and sees the couplings at the falling scale $\mu=1/L(\tau)$, $\tau\equiv ht$, as the torus inflates. %
Its survival probability on one branch is
\begin{equation}
P_1=\exp\Bigl(-\int\Gamma\,d\eta\Bigr),\qquad d\eta=\frac{d\tau}{hL(\tau)},
\label{eq:P1}
\end{equation}
where $\eta$ is the conformal time in units of the comoving edge and the branch lasts $\eta_\infty\approx300$; for the natural initial state at the turn the survival probability is $w_0P_1$, with $w_0=0.98$--$0.995$~\cite{v12PRD}. %
For $\xiH=1/6$ (see below) the cosmology enters only through $\mub$ and this integral, which equals that of a static box at $\mub$ lasting only $N_{\rm eff}=\NeffHalf$ light-crossing times, because $\Gamma$ decreases as $\lambda(\mu)$ rises along the branch [Fig.~\ref{fig:mechanism}(b)]. %
We find $\lamc(\mub)=-0.0121^{+\lamcErrUp}_{-\lamcErrDn}$ (model error, including the inflation scale), with $-0.0091$ and $-0.0152$ for $P_1=0.9$ and $0.1$ [Fig.~\ref{fig:mechanism}(b)]. %
For $m_t=172.60$~GeV and $\alpha_s(M_Z)=0.1180$ the SM value, $\lambda(\mub)=-0.0098$, lies on the survival side, $\marginLam$ above $\lamc$, or $\marginGeV$~GeV in the top mass; without the finite-volume threshold the criterion would be $\lambda(\mub)\geq0$, or $m_t<171.1$~GeV. %

\begin{table}[t]
\caption{Critical top mass $m_c$, at which $P_1=0.9$, $0.5$ and $0.1$, and its uncertainties in GeV, for $\alpha_s(M_Z)=0.1180$ and $M_h=125.13$~GeV; the last two rows give the uncertainties of the measured mass. %
Footnote-a items and the three model rows are combined in quadrature for each sign; model rows are for the pole mass and about 5\% smaller for $\bar m_c$.} %
\label{tab:budget}
\begin{ruledtabular}
\begin{tabular}{lccc}
 & $P_1=0.9$ & $P_1=0.5$ & $P_1=0.1$\\
\hline
$m_c$ (pole) & 172.50 & 172.96 & 173.42\\ %
$\bar m_c$ ($\MSbar$) & 162.73 & 163.16 & 163.60\\ %
\hline
Zero-mode model\footnotemark[1] & $^{+0.018}_{-0.040}$ & $^{+0.027}_{-0.051}$ & $^{+0.038}_{-0.061}$\\ %
Inflation scale\footnotemark[2] & $\pm0.035$ & $\pm0.067$ & $\pm0.11$\\ %
Preparation weight $w_0$ & $-0.016$ & $-0.006$ & $-0.004$\\ %
Model total & $^{+0.039}_{-0.056}$ & $^{+0.072}_{-0.084}$ & $^{+0.12}_{-0.13}$\\ %
SM mapping ($m_c$\,/\,$\bar m_c$) & \multicolumn{3}{c}{$\pm0.10$\,/\,$\pm0.07$}\\ %
$\alpha_s\pm0.0009$ ($m_c$\,/\,$\bar m_c$) & \multicolumn{3}{c}{$\pm0.35$\,/\,$\pm0.25$}\\ %
$M_h\pm0.11$~GeV & \multicolumn{3}{c}{$\pm0.05$}\\ %
\hline
Measured $m_t$ (direct) & \multicolumn{3}{c}{$\pm0.27$}\\ %
MC--pole interpretation & \multicolumn{3}{c}{$\pm0.5$}\\ %
\end{tabular}
\end{ruledtabular}
\footnotetext[1]{Reduction, one-loop matching, cross terms, full-dimensional check and numerics.}
\footnotetext[2]{Tensor-to-scalar ratio $\rT$ between $10^{-3}$ and $10^{-2}$ (plateau models; computed without the final corrections, which change it by at most 0.001~GeV).} %
\end{table}

\begin{figure}[t]
\centering
\includegraphics[width=\columnwidth]{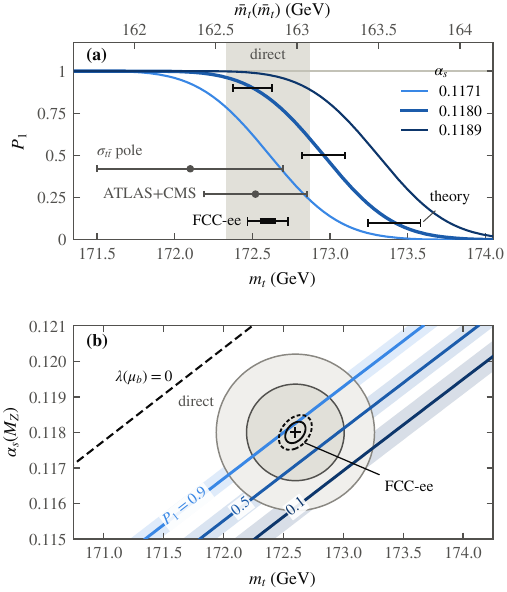}
\caption{Survival criterion in the top-mass plane.
(a)~$P_1(m_t)$ for $\alpha_s(M_Z)=0.1171$, $0.1180$ (thick) and $0.1189$, with theory $1\sigma$ bars; gray band: direct $m_t$; bars: $\sigma_{t\bar t}$ pole mass, ATLAS+CMS~\cite{ATLAS:2024dxp}, and FCC-ee (thick: $m_t$; thin: total); upper axis: $\bar m_t(\bar m_t)$ at $\alpha_s=0.1180$. %
(b)~Critical lines $P_1=0.9$, $0.5$, $0.1$ (theory $1\sigma$), $\lambda(\mub)=0$ (dashed; the absolute-stability bound lies just to its left), and 68\% and 95\% ellipses for present data and FCC-ee.}
\label{fig:criterion}
\end{figure}

\runin{Top-quark criterion}We map the measured SM parameters to $\lambda(\mub)$ with next-to-next-to-leading-order (NNLO) matching at the top mass and three-loop running, including the four- and five-loop QCD terms.
We use SMDR~\cite{Martin:2019lqd} and an implementation of our own; the spread with respect to mr~\cite{Kniehl:2016enc} is part of the SM error.
With $M_h=125.13$~GeV~\cite{ParticleDataGroup:2026mpi}, we obtain the criterion %
\begin{equation}
P_1>\tfrac12 \;\Leftrightarrow\; m_t<172.96~\mathrm{GeV}
\;\;\bigl[\bar m_t<163.16~\mathrm{GeV}\bigr], %
\label{eq:criterion}
\end{equation}
for $\alpha_s(M_Z)=0.1180$, with $m_t$ the pole mass obtained from $\bar m_t$ at four loops in QCD~\cite{Marquard:2015qpa}. %
The steep dependence on $m_t$ reflects the near-criticality of the SM~\cite{Degrassi:2012ry,Buttazzo:2013uya,Bednyakov:2015sca}: the top-Yukawa running that sets the absolute-stability bound gives $d\lambda(\mub)/dm_t\approx-0.0066~\mathrm{GeV}^{-1}$, which converts $\lamc$ into a top mass. %
The threshold places the criterion about $\mcMinusMzero$~GeV above the line $\lambda(\mub)=0$, which with our inputs lies just above the gauge-independent absolute-stability bound, $\min_\mu\lambda(\mu)=0$~\cite{Bednyakov:2015sca}, and far below the metastability bound, $177.7$~GeV~\cite{Andreassen:2017rzq} [Fig.~\ref{fig:criterion}(b)]. %

\runin{Current data and tests}The direct (Monte Carlo) mass, $172.60\pm0.27$~GeV~\cite{ParticleDataGroup:2026mpi}, which we identify with $m_t$ up to an interpretation uncertainty of about 0.5~GeV~\cite{Hoang:2020iah}, lies on the survival side: at its central value $P_1=0.84$ [Fig.~\ref{fig:criterion}(a)]. %
Marginalizing over $m_t$, $\alpha_s$, $M_h$ and the SM and model errors of Table~\ref{tab:budget}, we obtain $\bar P_1=\PbarOne$, or $\PbarInterp$ with the 0.5~GeV interpretation uncertainty. %
Other identifications of the measured mass give 0.64 (MSR mass) to $\PbarSigma$ (cross-section pole mass), and the $\bar m_t$ formulation reproduces the direct-mass value (End Matter B). %
The data do not yet discriminate: for every measurement and prior we considered, Bayes factors of the survival condition satisfy $|\ln B|\leq\BayesMax$. %
FCC-ee~\cite{FCC:2025lpp,Freitas:2019bre,Defranchis:2025auz} would separate $m_t=172.6$~GeV from $173.3$~GeV, where $P_1\approx0.17$, at $\sigFCC\sigma$, including the theory errors of the criterion. %
At 95\% confidence, a measured pole mass below $\mSurvFCC$~GeV would establish $P_1>1/2$, and one above $\mExclFCC$~GeV, 1.1~GeV above the present direct value, would establish $P_1<0.1$, with a Bayes factor $B_{\rm ex}\approx\BexFCC$ at the threshold (End Matter B). %
The HL-LHC would separate the same two masses at $\sigHLLHCmin$--$\sigHLLHCmax\sigma$ if the direct mass, measured to 0.17~GeV, is identified with the pole mass (limited by $\alpha_s$), and at $\sigHLLHCinterpMin$--$\sigHLLHCinterpMax\sigma$ with its interpretation uncertainty~\cite{Azzi:2019yne}. %

\begin{figure}[t]
\centering
\includegraphics[width=\columnwidth]{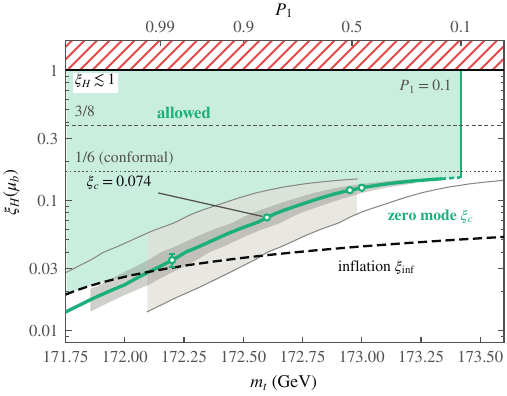}
\caption{Curvature coupling needed for survival.
The zero-mode threshold $\xi_c$ at $\mub$ and the bound from vacuum decay during plateau inflation (thick dashes) constrain $\xiH$ from below, and $\xiH\lesssim1$ from above. Solid: the universal curve $\xi_c(P_1)$ through the direct calculations (circles at $\alpha_s=0.1180$) and those of the SM and $\alpha_s$ variants, tested for $0.45\lesssim P_1\lesssim0.998$; short dashes: extrapolation below $P_1=0.13$; bands: SM mapping and $\alpha_s$. %
In the shaded region $P_1\geq0.1$ and survival is within 1\% of its conformal value.
Dotted: $\xiH=1/6$; thin dashed: $\xiH=3/8$, the onset of tachyonic resonance after inflation~\cite{Herranen:2015ima}.} %
\label{fig:xi}
\end{figure}

\runin{Curvature coupling}After the turn, the curvature term $\xiH R\,\Phi^\dagger\Phi$ controls the late part of the hazard.
It drops out of the zero-mode equation in the conformal frame for $\xiH=1/6$, to which $P_1$, $\bar P_1$ and the critical masses above refer. %
For $\xiH=0$ the zero mode begins to run away about four e-folds after the turn ($\tau\approx4$--5), and its survival tends to zero as inflation proceeds~\cite{v12PRD}. %
We follow the zero mode in real time into the classical regime and define $\xi_c$ as the coupling at $\mub$ for which survival is 1\% below its value at $\xiH=1/6$. %
Expressed through $P_1$, $\xi_c$ follows a single curve to within 0.002, so the SM and $\alpha_s$ errors enter only through $P_1$. %
At the measured top mass survival stays within 1\% of its conformal value for $\xiH(\mub)\gtrsim\xiHmin$ (central prescription $\xiHcentral$, about $\xiHminEW$ at the electroweak scale; SM and $\alpha_s$ errors $\pm\xiHbandSM$ and $\pm\xiHbandAs$), and this bound rises to $\xiHhalf$ at $P_1=1/2$ [Fig.~\ref{fig:xi}]; below it survival decreases gradually, by 3.5--5\% at $\xiH=0.04$, and at $\xiH=0.08$ the criterion moves by only $-0.01$~GeV. %
At this 1\% tolerance the zero-mode condition lies above the bound from vacuum decay during plateau inflation at the same scale~\cite{Mantziris:2020rzh}, $\xiH(\mub)\gtrsim0.038$ at $m_t=172.6$~GeV, recomputed with its method~\cite{v12PRD}. %
The stronger bound of Ref.~\cite{Mantziris:2022fuu} needs an $R^2$ term, absent here. %
The required coupling is not large: the conformal value $1/6$, a fixed point of the one-loop running of $\xiH$, satisfies every condition, and the tachyonic resonance after inflation destabilizes the Higgs only for $\xiH\gtrsim1$~\cite{Herranen:2015ima}. %

\runin{Discussion}The threshold is a property of the SM in a small box and rests on three premises: (i)~a flat three-torus without 't~Hooft flux whose fermions have no constant modes, with the one-loop vacuum at trivial holonomy; (ii)~SM couplings matched at $\mu=1/L$ at one loop; and (iii)~the reduced zero-mode model of End Matter A. The duration $N$ is an input, which a cosmology fixes through its expansion.
Because the resonance extends only to field values of about $6/L$, or $0.008\,\MPbar$ at $\mub$, a Planck-suppressed term $c\,r^6/\MPbar^2$ changes the effective quartic there by less than $3\times10^{-4}c$, and the critical top mass by about $0.04c$~GeV. %
An indicative estimate gives a threshold of the same size on tori with one or two periodic cycles, while in sectors with 't~Hooft flux the Higgs doublet has no constant mode and Landau levels replace the zero-point confinement, with $\lamc\approx-1$~\cite{v12Report}. %
The cosmological criterion adds (iv)~semiclassical Einstein gravity with a plateau inflaton that does not couple directly to the Higgs field; (v)~the reading of the turn as a beginning, with $P_1$ the survival probability of the resonance on each branch and $w_0$ that of the natural state; (vi)~the SM up to $2\pi/\Lb$, with NNLO matching and $M_h=125.13$~GeV; (vii)~$\xiH(\mub)\gtrsim\xic$; and (viii)~the top-mass interpretations of End Matter B. %
Within the SM, with its curvature coupling $\xiH$, the criterion requires no new physics in the Higgs sector.
The joint evolution of the zero and nonzero Higgs modes after the torus leaves the Hubble radius is not simulated; because superhorizon nonzero modes grow only briefly inside the curvature barrier, it acts only on the few percent of the probability near the barrier at $\xiH\approx\xic$ and can only raise $\xic$~\cite{v12PRD}. %

\runin{Conclusion}Zero-point fluctuations of the constant gauge modes turn the runaway of a negative Higgs quartic coupling in a small box into a resonance with a finite threshold $\lamc$, set by the gauge and top Yukawa couplings and, through a logarithm only, by the lifetime of the box.
A universe that begins at a Casimir turning point probes this threshold at $\mub$, where $\lamc=-0.0121$ and the SM value clears it for $m_t<172.96$~GeV ($\bar m_t<163.16$~GeV); the electroweak vacuum then survives on each branch with probability $\bar P_1\approx\PbarOneAbs$ for current data, given $\xiH(\mub)\gtrsim\xiHmin$. %
The total model error of the criterion, $+\modelErrUp/{-\modelErrDn}$~GeV, is smaller than the measurement errors of $m_t$ and $\alpha_s$. %
A measured top-quark pole mass above $\mExclFCC$~GeV, within reach of FCC-ee, would establish $P_1<0.1$ at $95\%$ confidence, with a Bayes factor of about 30 against this beginning (End Matter B). %

\begin{acknowledgments}
\runin{Acknowledgments}The author acknowledges graduate scholarship and research support through McMaster University.
Research at Perimeter Institute is supported in part by the Government of Canada through the Department of Innovation, Science and Economic Development and by the Province of Ontario through the Ministry of Colleges, Universities, Research Excellence and Security.
Parts of the derivations and of the numerical code were developed with Anthropic Claude Code (model Claude Opus~5.5) and OpenAI Codex (model gpt-6-astra); separately instructed instances of these tools, without access to the author's derivations or code, carried out the reimplementations described in End Matter A.
The same tools were used for literature screening, manuscript drafting and English-language editing.
The author checked all derivations, numbers and conclusions and takes full responsibility for this work.
\end{acknowledgments}

\runin{Data availability}The code and numerical data that support the findings of this article are openly available~\cite{DNECv12code}; a development version is maintained at \url{https://github.com/Ailun-Zhang/casimir-cosmology}. The code and data of the static-box and abelian-Higgs calculations (Fig.~\ref{fig:mechanism}, End Matter A) are included in the Supplemental Material~\cite{PRLsupp}.

\bibliography{refs}

\section*{End Matter}
\appendix
\setcounter{equation}{0}
\renewcommand{\theequation}{A\arabic{equation}}
\renewcommand{\theHequation}{A\arabic{equation}}

\runin{Appendix A: The zero-mode model, the static box and the hazard}We take the Higgs zero mode in Coulomb gauge, with flat measure and canonical kinetic term, and impose Gauss's law for constant gauge transformations on Eq.~\eqref{eq:H}.
In a body frame aligned with $\Phi$, the SU(2) zero modes are taken in the most general invariant centered Gaussian conditional on $r$ and $\mathbf w$; for this family the factorization is exact, and its Born--Oppenheimer terms are kept.
The radius $r$ and the hypercharge modes are then solved together with
\begin{multline}
H_{\rm 2D}=-\tfrac12\partial_r^2+\frac{3}{8r^2}-\tfrac12\Delta_{\mathbf w}+C_0(r)+B_x(r)+\tfrac12\Omega_{\rm eff}^2(r)\,\mathbf w^2\\
+E_X(r)+\tfrac12\Omega_X^2(r)\,\mathbf w^2+\tfrac{\lambda}{4}r^4+\Delta V(r)+\delta E_{\rm DMC}(r),
\label{eq:H2D}
\end{multline}
where $C_0+\tfrac12\Omega^2\mathbf w^2$ is the conditional SU(2) energy, $\Omega_{\rm eff}^2$ adds to $\Omega^2$ the Born--Oppenheimer term of the $r$-dependent shift of the SU(2) Gaussian, $B_x$ is the remaining Born--Oppenheimer term, $E_X+\tfrac12\Omega_X^2\mathbf w^2$ are the cross terms, and $\delta E_{\rm DMC}$ is the difference between DMC~\cite{Foulkes:2001zz} and the Gaussian in the SU(2) sector.
For each field of mass $m(r)$ the matching potential $\Delta V$ contains the finite-volume sum over nonzero modes and the infinite-volume Coleman--Weinberg term,
\begin{equation}
E(m^2)=D_\theta(m^2)+\frac{m^4}{64\pi^2}\Bigl(\ln\frac{|m^2|}{\mu^2}-c\Bigr),
\label{eq:matching}
\end{equation}
with $D_\theta(m^2)=-\frac{1}{4\sqrt\pi}\int_0^\infty ds\,s^{-3/2}(e^{-m^2s}-1)\bigl[K_\theta(s)-\delta_{\theta0}-(4\pi s)^{-3/2}\bigr]$, $K_\theta(s)=\sum_{\mathbf n}e^{-4\pi^2|\mathbf n+\theta|^2s}$, $\theta=0$ for bosons and $(\frac12,\frac12,\frac12)$ for fermions, and $c=3/2$ except for transverse vector modes, which have $c=1/2$.
The logarithms cancel between the two terms, so $E$ is analytic in $m^2$ for $|m|<2\pi/L$ ($\sqrt3\pi/L$ for the antiperiodic fermions), and for $m^2<0$ the integral for $D_\theta$ is continued analytically.
Each massive vector contributes two transverse modes and a Coulomb mode that mixes with the Goldstone boson, and the top quark enters with weight $-12$.
With $\mu=1/L$ and the Coulomb-gauge anomalous dimension of $\Phi$, the scale dependence of $\Delta V$ matches the one-loop running of $\lambda$~\cite{v12PRD}.
We estimate the two-loop remainder by varying the matching scale from $1/(2L)$ to $2\pi/L$ and the scale of the top Yukawa coupling in the top-quark loop by a factor of two, which moves the critical top mass by $+0.01/{-0.05}$~GeV; in a simpler one-dimensional version of the model, a Landau-gauge matching moved it by $-0.012$ to $-0.016$~GeV~\cite{v12Report}. %
$\Delta V$ is evaluated at zero holonomy; its holonomy dependence gives the cross terms.
The resonance $E=\varepsilon-i\Gamma/2$ of Eq.~\eqref{eq:H2D} is the zero of a multichannel outgoing-wave (Siegert) condition~\cite{Siegert:1939zz}.
A complex absorbing potential~\cite{Riss:1993cap} reproduces the Siegert width of the final model to 0.08\%. %

For the static box of Fig.~\ref{fig:mechanism} the couplings are fixed at their SM values at $\mu=1/L$ for $m_t=172.60$~GeV and $\alpha_s(M_Z)=0.1180$, and $\lambda$ is varied; the widths come from Eq.~\eqref{eq:H2D} with the cross terms, and $\lamc(N)$ from $\Gamma N=\ln2$ by interpolation of $\ln\Gamma$ in $|\lambda|^{-1/2}$. %
With the couplings of the expanding torus the same code reproduces the archived widths of the final model exactly, and to $\SzeroGeneric$ from generic starting points. %
A narrow resonance, with $\varepsilon$ below the barrier top, exists for $\lambda\geq\lamTop$ at $\mu=\mub$, so that the threshold for $P=1/2$ is defined for $N\gtrsim10$. %
The model error of $\lamc(300)$ is the zero-mode-model row of Table~\ref{tab:budget} converted with $d\lambda(\mub)/dm_t$; the static box and the expanding torus share it, as their cross-term shifts, $0.00075$ and $0.00078$, show. %
For the $\kappa$ scan, $g_2$ and $g_Y$ are multiplied by $\kappa$ in $C_0$, $B_x$, $\Omega_{\rm eff}$ and the hypercharge oscillator, and the DMC correction is carried along with the exact zero-mode scaling, $\delta E_{\rm DMC}\to\kappa^{2/3}\delta E_{\rm DMC}(\kappa^{1/3}r)$; $\Delta V$ stays at the physical couplings, and the cross terms are left out. %
The scan is a diagnostic, since for $\kappa\neq1$ this split does not follow from one Lagrangian; below $\kappa\approx\kappaMin$ no narrow resonance exists even at $\lambda=0$. %
In the abelian Higgs model, with charge $e$ and neither fermion constant modes nor matching, Gauss's law and rotation invariance in $\mathbf w$ reduce the zero-mode problem exactly to $r$ and $|\mathbf w|$.
Its resonance, from exterior complex scaling checked with a complex absorbing potential to $\abelCAP$, has a local exponent in $(e^2/|\lambda|)^{1/2}$ that rises from $\abelSlopeLo$ (level above the barrier) to $\abelSlopeHi$ as $\lambda/e^2$ goes from $\abelLamMin$ to $\abelLamMax$; the finite resonance energy leaves a deficit $\propto|\lambda|^{1/3}$ below the value $\abelSlopeWKB$ of Eq.~\eqref{eq:Gamma} with $\sigma=\sigma_0=3e/2$. %
Plain Born--Oppenheimer widths are within $\abelBOdev$, whereas adding the diagonal Born--Oppenheimer term overestimates them by factors of 2.5 to more than 200; our reduction does not use that term for $\mathbf w$, since it solves $r$ and $|\mathbf w|$ together. %
The model tests only the mechanism: at $\sigma_0=1.31$ ($e=0.87$) its $\lamc(300)=\abelLamcSM$ is six times deeper than in the SM, whose $\Delta V$ and SU(2) zero-mode energy it lacks, and its non-compact treatment of the holonomy affects a probability below $\abelHolonomy$. %

The hazard of the expanding torus uses eleven nodes $\tau=0,0.5,\ldots,5$ with SM couplings at $\mu=\mub(\cosh2\tau)^{-1/2}$, and integrates a cubic spline of $\ln\Gamma$ in Eq.~\eqref{eq:P1}; at later times the zero mode is followed in real time. %
Because $\lambda(\mu)$ rises as $\mu$ falls, $\Gamma$ decreases along the branch, and the static threshold at $\mub$ for $N=\eta_\infty$ is shallower than the hazard threshold by $\staticHazardDiff$. %
Full-dimensional DMC with a Dirichlet wall at fixed $r$, the only exact node, shifts $\varepsilon$ by $-0.003$ ($-0.004$ at $\lambda=0$) relative to the reduced model with the same wall, independently of the wall and of $\lambda$, at fixed couplings without cross terms. %
With WKB estimates, tail amplitudes ($0.92\pm0.16$ and $1.03\pm0.07$) and three-dimensional collective coordinates this gives $\Gamma_{\rm full}/\Gamma_{\rm 2D}=0.92$--$1.03$, with 0.976 from the energy shift and 0.999 from the collective coordinates; Table~\ref{tab:budget} carries the WKB value, 0.94, which lies below both, as a one-sided item of at most $+0.02$~GeV ($+0.03$~GeV at $P_1=0.1$) that the central values do not include. %
The cross terms follow from the image sums of all SM fields on the antiperiodic torus, twisted by the holonomies~\cite{v12PRD}.
Their shift of the critical mass at $P_1=1/2$, $-0.12$~GeV, splits into $-0.08$~GeV from the electroweak zero modes, mainly through the hypercharge curvature of the antiperiodic light fermions, and $-0.04$~GeV from the gluon zero modes through the top quark. %
A fixed-coupling variational estimate and a blind reimplementation agree on this split~\cite{v12PRD}. %

The reimplementations also reproduce the SM mapping and the marginalization; they test the implementation, while Table~\ref{tab:budget} covers the approximations. %
Three parts of the calculation were checked only within our own code: refinements of the cross terms that move the critical masses by less than 0.01~GeV, the final matching-scale scan, and the real-time calculation of $\xi_c$. %
Symbolic steps were checked in SymPy or Wolfram Language. Every number quoted here is reproduced by the scripts of Ref.~\cite{DNECv12code} (for the static-box and abelian calculations, by those of the Supplemental Material~\cite{PRLsupp}) or contained in their archived outputs, including those of the blind reimplementations, as documented in Ref.~\cite{v12Report}.

\setcounter{equation}{0}
\renewcommand{\theequation}{B\arabic{equation}}
\renewcommand{\theHequation}{B\arabic{equation}}

\runin{Appendix B: Top-mass definitions, marginalization and reach}The pole mass $m_t$ follows from $\bar m_t\equiv\bar m_t(\bar m_t)$ through the four-loop QCD relation~\cite{Marquard:2015qpa}, truncated at that order; the all-orders pole mass would raise the criterion by 0.27~GeV, and the renormalon ambiguity of the pole mass is estimated at 0.07--0.25~GeV~\cite{Beneke:2016cbu,Hoang:2017btd}. %
The short-distance mass $\bar m_t$ carries neither, which is why Eq.~\eqref{eq:criterion} quotes both.
We consider four identifications of the measurements: the direct mass, $172.60\pm0.27$~GeV, as $m_t$, with or without an interpretation uncertainty of 0.5~GeV~\cite{Hoang:2020iah,Butenschoen:2016lpz}; the direct mass as the MSR mass at $R=1.3$~GeV, with $\pm0.5$~GeV (172.96~GeV in the pole mass at the order of turnover, 172.69~GeV in the four-loop pole mass used here); the cross-section pole mass, $172.1\pm0.6$~GeV~\cite{ParticleDataGroup:2026mpi}; and the direct mass converted to $\bar m_t$. %
The marginalized probability is
\begin{equation}
\bar P_1=\int P_1(m_t,\alpha_s,M_h;\delta_{\rm SM},\delta_{\rm model})\,dP,
\label{eq:Pbar}
\end{equation}
with Gaussian $m_t$, $\alpha_s=0.1180\pm0.0009$~\cite{ParticleDataGroup:2026mpi}, $M_h=125.13\pm0.11$~GeV, SM error $\pm0.10$~GeV and the model errors of Table~\ref{tab:budget}; asymmetric errors are split into the two sides with equal probability. %
This gives $\bar P_1=\PbarOne$ for the direct mass, $\PbarInterp$ with the interpretation uncertainty, 0.64 for the MSR mass, $\PbarSigma$ for the cross-section pole mass, and the direct value again in the $\bar m_t$ formulation; the BKPV/FJ organization of the two-loop matching~\cite{Bednyakov:2015sca} gives $\PbarBKPV$. %
Large same-sign non-QCD logarithms of $\lambda$ and $y_t$ in its $G_F$ scheme widen its scale variation to $+0.17/{-0.36}$~GeV in $m_c$, so we keep it as a variant. %
The inflation-scale row of Table~\ref{tab:budget} covers tensor-to-scalar ratios between $10^{-3}$ and $10^{-2}$; for $\rT=10^{-4}$ the critical mass at $P_1=1/2$ would be lower by 0.17~GeV. %
Halving or doubling the duration of the turn changes $|\lamc|$ by $+13\%$ or $-11\%$. %
The marginalized probability is itself the Bayes factor $B_{\rm ex}=\bar P_1$ of the hypothesis that the zero mode survived against the same hypothesis without that requirement~\cite{v12PRD}; we call this beginning excluded when a measured $m_t$ establishes $P_1<0.1$ at 95\% confidence, and $B_{\rm ex}$ measures the strength of that exclusion. %
For FCC-ee we assume $\delta\bar m_t=50$~MeV from the threshold scan and $\delta\alpha_s=0.0002$~\cite{FCC:2025lpp,Freitas:2019bre,Defranchis:2025auz}, $\delta M_h=0.01$~GeV, our SM error of 0.07~GeV and the model errors of Table~\ref{tab:budget}; with $\delta\bar m_t=38$~MeV and $\delta\alpha_s=0.0001$ the separation of 172.6 and 173.3~GeV grows to $\sigFCCamb\sigma$. %
For the baseline inputs, at the exclusion threshold the central $P_1$ is $\approx\PoneAtExclFCC$ and the Bayes factor $B_{\rm ex}\approx\BexFCC$; above the threshold both are smaller. %
For the HL-LHC we take $\delta m_t=0.17$~GeV and $\delta M_h=0.03$~GeV~\cite{Azzi:2019yne}, with $\delta\alpha_s=0.0009$, or $0.00045$ if lattice determinations improve twofold~\cite{FCC:2025lpp}. %
Both branches of the time-symmetric history carry the same $P_1$.
Their joint survival has no decoherent-histories probability in this model; the only time-symmetric linear value, $P_2=P_1^2/w_0$, would lower the critical mass for probability $1/2$ ($0.9$) by 0.20 (0.12)~GeV~\cite{v12PRD}. %

\end{document}